\documentclass[prl,twocolumn,aps]{revtex4}
\usepackage{verbatim}
\usepackage{graphicx}
\normalfont
\begin{document}


\title{Overhauser effect in individual InP/GaInP dots}

\author{J. Skiba-Szymanska$^{1}$, E. A. Chekhovich$^{1}$, A. E. Nikolaenko$^1$, A. I. Tartakovskii$^1$, M. N. Makhonin$^{1}$, I. Drouzas$^1$, M. S. Skolnick$^1$, A. B. Krysa$^3$}
\address{$^{1}$ Department of Physics and Astronomy, University of Sheffield, S3 7RH,UK \\ $^{3}$ Department of Electronic and Electrical Engineering, University of Sheffield, Sheffield S1 3JD, UK}

\date{\today}

\begin{abstract}

Sizable nuclear spin polarization is pumped in individual electron-charged InP/GaInP dots  in a wide range of external magnetic fields  $B_{Z}=0-5$T by circularly polarized optical excitation. We observe nuclear polarization of up to $\approx40\%$ at $B_{Z}=1.5$T  corresponding to an Overhauser field of $\approx1.2$T. We find a strong feedback of the nuclear spin on the spin pumping efficiency. This feedback, produced by the Overhauser field, leads to nuclear spin bi-stability at low magnetic fields of $B_{Z}\approx 0.3-1$T. We find that the splitting in magnetic field between the trion radiative recombination peaks increases markedly, when the Overhauser field in the dot cancels the external field. This counter-intuitive result is shown to arise from the opposite contribution of the electron and hole Zeeman splittings to the transition energies.\\
\end{abstract}

\maketitle


{\bf I. INTRODUCTION}

Recent progress in nano-science and technology has allowed access to desirable properties of single electron and hole spin states in semiconductor nano-structures \cite{Koppens,Petta,Atature,Xu}, that can be addressed both optically \cite{Atature,Xu} and electrically \cite{Koppens,Petta}. It has been demonstrated that due to suppression of the spin-orbit interaction in quantum dots, $T_1$ of the electron spin is in the ms range \cite{Kroutvar}, opening potential applications in quantum information processing. Of particular importance in this context is the electron-nuclear spin interaction in quantum dots, representing a major source of decoherence of electron-spin based qubits \cite{Khaetskii,Petta}.

Several approaches to overcome such decoherence have been suggested mainly focused on nuclear spin cooling methods \cite{Klauser}. Dynamic nuclear polarization arising under circularly polarized optical excitation so far resulted in degrees of nuclear polarization $S_N\approx 60\%$ and $50\%$ for interface (GaAs/AlGaAs \cite{Gammon}) and self-assembled (InGaAs/GaAs \cite{Urbaszek}) GaAs-based dots, respectively. The reasons for the relatively low degrees of nuclear polarization are largely unclear. Nuclear spin pumping relies on the electron-nuclear spin flip-flop and may be slowed down due to the large electron Zeeman splitting either due to the external ($B_{Z}$)\cite{Braun2,Maletinsky1,Tartakovskii} or nuclear (Overhauser, $B_N$) magnetic field\cite{Oulton}.  The pumping competes with nuclear spin diffusion into the matrix outside the dot \cite{Makhonin}, which may prevent high nuclear polarization degrees. Slowing down of the spin cooling rate is also possible due to formation of ''dark'' nuclear states \cite{Imamoglu}. 

Currently available III-V semiconductor QDs offer access to small isolated ensembles of nuclear isotopes with spins ranging from 1/2 (for P$^{31}$) to 9/2 (for In$^{115}$). New insights into the electron-nuclear spin interactions and nuclear spin cooling in particular are possible from the study of different types of QDs, where the whole nuclear spin ensemble as well as each individual nucleus experience different magnetic surrounding. In this work we study a III-V InP/GaInP quantum dot system \cite{Masumoto2,Pal,Dzhioev2,Kozin,Ikezawa,Sugisaki}, which, compared to a well-studied GaAs-based dots, provide electron spin states with a large g-factor \cite{Masumoto2}, and a possibility, in principle, to manipulate phosphorus nuclei possessing a simple spin configuration with $I_P=1/2$. In contrast, in (In)GaAs dots all isotopes possess nuclear spin $I\geq3/2$ and more complex nuclear spin pumping mechanisms may take place.

This work reports on the nuclear spin pumping in an individual electron-doped dot in a III-V system. This opens up possibilities to study the influence of the hyperfine interaction on the optically controlled electron spin with the life-time not limited by interaction with the electron reservoir in the contact \cite{Maletinsky2,Smith} or fast electron-hole recombination \cite{Tartakovskii,Russell}. Strong implications for the nuclear spin dynamics in the presence of the resident electron are also expected \cite{Deng}.  

More specifically, this paper reports on optically induced Overhauser fields up to 1.2 Tesla in individual InP/GaInP dots charged with a single electron in a wide range of external fields $B_{Z}=0-5$T. A strong dependence of the spin pumping efficiency on the circular polarization of the incident light is found, a manifestation of strong feedback of the optically pumped nuclear spin on the electron-to-nuclei spin transfer efficiency. The highest degree of nuclear polarization in an InP dot is $S_N^{max}\approx40\%$ (at $B_{Z}=1.5$T). We find that the splitting in magnetic field between the trion recombination peaks in a dot markedly increases under the conditions of positive feedback, when the Overhauser field $B_N$ is anti-parallel to the external field and the electron Zeeman splitting is minimized. We show that this initially counter-intuitive increase of the  total splitting, $E_{xZ}$, is the consequence of the  opposite contributions to $E_{xZ}$ of the (smaller) electron and (larger) hole Zeeman splittings. We also find that the feedback of the nuclear field $B_N$ on the electron-nuclear spin transfer rate results in  nuclear spin bi-stability, which we observe in the range of $B_{Z}\approx 0.3-1$T. \\

\begin{figure}
\centering
\includegraphics[height=7cm]{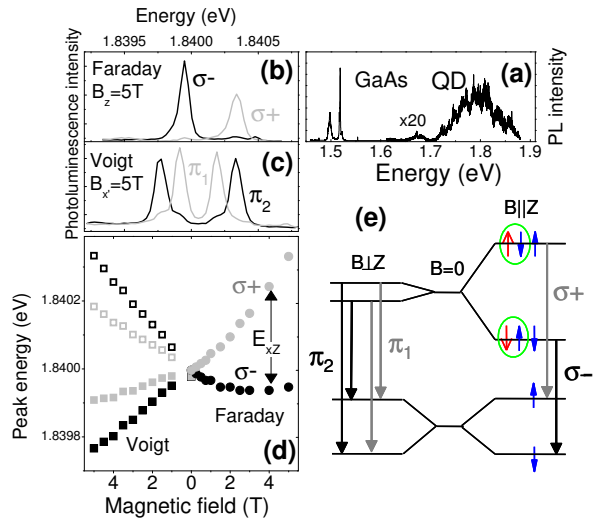}
\caption{(a) A typical PL spectrum of ensemble of InP/GaInP dot measured at $T=15$K. Emission from the GaAs substrate is observed around 1.51eV. (b) Trion PL spectra measured at $B_Z=5$T for a single InP/GaInP dot using linearly polarized excitation and $\sigma^+$ (gray) and $\sigma^-$ (black) detection. (c) Trion PL spectra measured at in-plane field $B_{X'}=5$T for a single InP/GaInP dot using linearly polarized detection. (d) Trion peak energies versus external magnetic field in the Voigt anf Faraday configurations. (c)  Schematic diagram of allowed optical transitions in a negatively charged InP dot in Voigt and Faraday configurations. Electrons (holes) are shown with small (large) arrows schematically representing spin-up and spin-down states. After emission of a $\sigma^+$($\sigma^-$)-polarized photon, an electron with spin up (down) is left on the dot.}
\label{fig1}
\end{figure}

{\bf II. SAMPLE AND EXPERIMENTAL METHODS}

The InP dots in the GaInP matrix studied in this work were grown by low-pressure metalorganic vapour phase epitaxy in a horizontal flow quartz reactor. The samples were grown on (100) GaAs substrates with a 10$^o$ misorientation towards $<$111$>$, used to suppress the CuPt-type ordering in the GaInP matrix. The growth temperature of the GaAs buffer and bottom GaInP layer was 690C. Before the deposition of InP, the wafer was cooled to 650C. After the deposition of InP and formation of the dot layer, the growth temperature was again raised to 690C, and a GaInP capping layer was deposited without growth interruption. The grown GaInP layers were nearly (within 0.04$\%$) lattice matched to GaAs as derived from X-ray diffractometry measurements. The growth rates for the GaAs and GaInP layers were $\approx$0.7 nm/s and for InP $\approx$0.35 nm/s \cite{Lewis}. For such growth conditions, InP dots with a density of $\approx 10^{10}$cm$^{-2}$ are formed. A typical low temperature (15K) photoluminescence (PL) spectrum of an ensemble of InP/GaInP dots excited with a HeNe laser at 633 nm is shown in Fig.1a. The dot PL is centered around 1.79eV with a full width at half maximum (FWHM) of the peak of 100meV. 

The as grown wafer was then covered with a 10/90 nm Ti/Al shadow mask, with 400 nm diameter clear apertures fabricated by means of electron beam lithography to allow optical access to individual QDs. PL was excited with a semiconductor diode laser emitting at 650 nm. A standard micro-PL set-up was employed, with the sample mounted on a cold finger (at temperature 15K) in a continuous flow helium cryostat equipped with a superconducting magnet. Both Faraday and Voigt geometries were employed and PL was measured with a double spectrometer and a liquid nitrogen cooled CCD camera.\\

{\bf III. PL CHARACTERIZATION}

At zero magnetic field a typical PL spectrum of an individual InP/GaInP QD in our sample consists of a single line exhibiting no fine structure splitting, a signature of dot charging \cite{Finley,Bayer,Tartakovskii1}. Under excitation with circularly polarized light, PL exhibits negative circular polarization, the degree of which increases with excitation power and can reach up to $30\%$. This effect is observed at zero field and in a wide range of magnetic fields $B_Z$ applied in the growth direction. Such behavior was previously found in both InGaAs and InP dots charged with a single electron and corresponds to optical orientation of the spin of the resident electron left behind after recombination of the optically excited trion \cite{Cortez,Oulton,Pal}. It has been found that in negatively charged dots excitation with $\sigma^+$ ($\sigma^-$) leads to stronger PL in $\sigma^-$ ($\sigma^+$) polarization leaving a localized electron with predominantly spin down (up).  

Additional evidence for the dot charging is obtained from PL measurements in magnetic field. Fig.1b shows typical exciton PL spectra of an individual InP/GaInP QD excited with linearly polarized light in magnetic field $B_Z=5T$ along the sample growth direction. A Zeeman doublet is measured with high and low energy peaks observed in $\sigma^+$ and $\sigma^-$ circular polarization, respectively. In the field applied in the plane of the dot $B_{X'}$, the emission line splits into four linearly polarized peaks with co-polarized inner and outer pairs (Fig.1c). 

Fig.1d shows a summary of peak positions measured in magnetic fields 0-5T applied either in the growth direction (circles) or in-plane (squares). In the Voigt  geometry (with the in-plane field) all four lines exhibit nearly linear energy shifts with a very small diamagnetic component and a common origin at the spectral position of the line at $B=0$. The behavior observed in Fig.1d, found previously for singly-charged dots \cite{Tischler,Akimov}, is in striking contrast to what expected for a neutral exciton, PL spectra of which in the Voigt geometry may have up to four lines originating from the pairs of dark and bright exciton states split at $B=0$ by the electron-hole exchange interaction \cite{Tischler,Bayer}. Bright neutral exciton states at $B=0$ are also split by the electron-hole exchange interaction usually spectrally resolved in PL \cite{Finley,Bayer,Kulakovskii,Ebbens}.

Identifications of the four lines in Fig.1c,d is conducted by comparison with the diagram in Fig.1e where the scheme of optical transitions of a negatively charged exciton in Voigt (Faraday) geometry is shown in the left (right) part of the figure. The four lines in the Voigt configuration in Fig.1c originate from the hole spin splitting in the initial state and the electron spin splitting in the final state. The splittings between the four lines in Fig.1c,d are found to depend on the direction of the in-plane magnetic field. Such dependence, a signature of a low in-plane symmetry of the dot, originates from the variation of the hole g-factor, whereas the electron g-factor is expected to be isotropic.  Based on this consideration and comparing results obtained for various in-plane directions of B-field we deduce that in Fig.1d $g_{hX'}=0.5$ and  $g_{e}=1.46$, the latter with high accuracy being the same in other in-plane directions. We assume that this magnitude of the electron g-factor can also be used for the experiments in the Faraday geometry.    

In the Faraday geometry two peaks are observed that exhibit the Zeeman splitting and notable diamagnetic shifts (circles in Fig.1d). The peak splitting in the Faraday geometry in Fig.1b,d is well described by the expression $E_{xZ}=g_{x}\mu_BB_{Z}$ (see the vertical arrow in Fig.1d), where $g_x=1.35$ is an effective g-factor describing the splitting between the trion PL peaks. $g_x$ is smaller than $g_e$ indicating that $g_e$ and $g_h$ have {\it opposite} contributions to the resulting magnitude of the trion peak splitting. Since $g_h$ is expected to be larger than the in-plane g-factor $g_{hX'}$ \cite{Tischler,Bayer} we conclude that $|g_h|>|g_e|$ and $|g_x|=|g_h|-|g_e|$ with $|g_h|\approx 2.8$. In order to obtain  full agreement with experiment (Fig.1b), where the line in $\sigma^{+}$($\sigma^-$) is observed at high (low) energy, both electron and hole should have positive g-factors as depicted in the right part of Fig.1e. 

The majority of spectrally isolated PL lines in our sample showed the properties described above. Based on the evidence presented we assume that the majority of dots in the sample studied in this work are electron-charged, probably due to a low level residual doping in the bulk material. Although hole-charging will produce similar patterns of peaks in magnetic field, negative circular polarization has never been observed for positively charged trions.  Nuclear spin effects in the dots studied in this work (discussed below) further confirm our conclusions about the electron charging of the dots and the relation between the electron and hole g-factors. \\

{\bf IV. NUCLEAR SPIN PUMPING}

The electron-nuclear hyperfine spin interaction leads to a finite probability of spin exchange (spin ''flip-flop'') between the resident electron confined in the dot and a single nucleus of the large (about $10^4$) ensemble of nuclei.  Re-pumping of the spin-polarized electron on the dot occurring under circularly polarized optical excitation leads to a build-up of sizable nuclear spin polarization on the dot, $S_N$. The nuclear spin pumping efficiency can be described  by the probability of the electron-nuclear spin flip-flop. The efficiency of this process decreases with the increasing electron Zeeman splitting $E_e$ which is the major energy cost of the spin flip-flop \cite{Lai,Eble,Urbaszek,Braun2,Maletinsky1,book,Russell}. 

The collective effect of all nuclei on the dot can be described in terms of the occurrence of local nuclear magnetic fields $B_N \propto S_N$ leading to modification of the electron Zeeman splitting $E_{e}=|g_e|\mu_B(B_{Z}\pm B_N)$, the effect of nuclei on the hole splitting being negligible. The Overhauser shifts $\delta E=\pm|g_e|\mu_BB_N$  can be evidenced in PL experiments on individual dots as a modification of the trion splitting. The opposing contribution of the hole and electron Zeeman splittings to the observed trion peak splitting deduced above implies that the dynamic nuclear polarization in external field along $Z$-direction will result in modification of $E_{xZ}$ in the following way:  
\begin{equation}
E_{xZ}=|g_h|\mu_BB_{Z}-|g_e|\mu_B (B_{Z}\pm B_N).
\label{ExZ}
\end{equation}
In addition, as discussed above, the spin pumping efficiency is strongly dependent on $E_{e}$, and is therefore sensitive to $B_N$.  

\begin{figure}
\centering
\includegraphics[height=5.5cm]{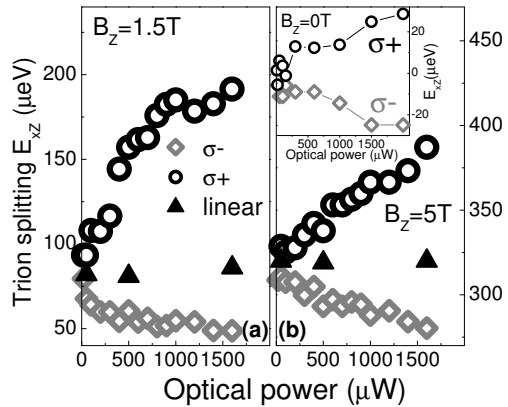}
\caption{Power dependences of the splitting $E_{xZ}$ (see Fig.1d) measured for a negatively charged exciton at $B_{Z}=1.5$ (a) and 5T (b) for $\sigma^-$ (diamonds), $\sigma^+$ (circles) and linearly (triangles) polarized excitation. The inset in (b) shows $E_{xZ}$ power dependence measured at $B_{Z}=0$T.}
\label{fig2}
\end{figure}

Fig.2 shows the power dependence of the splitting $E_{xZ}$ measured for a single dot trion at $B_{Z}=1.5$ and 5T for $\sigma^+$ and $\sigma^-$ circularly polarized excitation. Data for a different dot to that described in Fig.1 is reported in Fig.2. In Fig.2a the splitting changes from 85 $\mu$eV at small powers to 191(47) $\mu$eV at high powers for $\sigma^+$($\sigma^-$) excitation. The modification of $E_{xZ}$ is related to the pumping of the nuclear spin on the dot due to spin exchange with the resident spin-polarized electrons. The pumping, being a dynamical process competing with nuclear spin depolarization \cite{Maletinsky2,Makhonin}, becomes more efficient at higher powers as the rate of excitation of the electron spin on the dot increases. Clearly, much more efficient pumping is observed in the case of $\sigma^+$ excitation: the total splitting changes before saturation in Fig.2a are +106 and -38 $\mu$eV for $\sigma^+$ and $\sigma^-$ excitation, respectively. A negligible power dependence is observed for linearly polarized excitation (triangles in Fig.2) with the splitting between the lines close to that observed for low power pumping with circularly polarized light.   

An important feature is observed in Fig.2a: the more efficient nuclear spin pumping is achieved in the case where $E_{xZ}$ increases. The explanation can be found if Eq.\ref{ExZ} is considered. For $\sigma^+$ excitation $B_N$ anti-parallel to $B_{Z}$ is expected and $E_{e}$ is strongly reduced. Similar effect is observed in electron charged InGaAs dots \cite{Eble,Lai}, where nuclear spin pumping occurs due to the spin relaxation of the extra electron tunneled into the dot from the contact. As observed in Fig.2a for $\sigma^+$ exscitation, the total trion line splitting increases. On the other hand, the reduction of $E_{e}$ observed for $\sigma^+$ excitation will result in a positive feedback on the nuclear spin pumping efficiency. A negative feedback and as a consequence, a slower rate of the spin pumping is observed for $\sigma^-$ excitation, for which $E_{e}$ increases leading to a decrease of $E_{xZ}$ as predicted by Eq.\ref{ExZ}. 

The manifestation of the positive feedback in InP dots, i.e. more efficient nuclear spin pumping leading to the {\it increase} of the trion line splitting $E_{xZ}$, is opposite to that observed for InGaAs dots \cite{Braun2,Maletinsky1,Tartakovskii}. However, this difference is explained by the different relation between the electron and hole g-factors in the two types of dots: $g_e$ and $g_h$ have contributions of the same and opposite signs to the exciton Zeeman splitting for InGaAs/GaAs and InP/GaInP dots, respectively. The observations in both cases are consistent in that the positive feedback occurs due to the decrease of $E_{e}$ when $B_N$ is anti-parallel to $B_{Z}$.

From the data in Fig.2a, we obtain maximum $B_N\approx 1.2$T for $\sigma^+$ excitation (with $B_N$ anti-parallel to $B_{Z}$) and $B_N\approx 0.4$T for $\sigma^-$ excitation (with $B_N$ parallel to $B_{Z}$). The high spin pumping efficiency in the case of $\sigma^+$ excitation in Fig.2a can thus be explained by almost complete compensation of  $B_{Z}$ by $B_N$, resulting in negligible $E_{e}$ and high probability of the flip-flop. Note, that significantly larger Overhauser fields compared to previously found for InP dots \cite{Dzhioev2,Pal} are reported here.

We now estimate the degree of nuclear spin polarization on the dot. For this we assume that the dot contains In and P nuclei only with hyperfine constants $A_{In}=56\mu$eV and $A_{P}=44\mu$eV\cite{Gotschy}. The spins of In and P nuclei are $I_{In}=9/2$ and $I_{P}=1/2$, respectively. Fully polarized material will then produce an Overhauser shift of $\Sigma_{OH}=I_{In}A_{In}+I_{P}A_{P}=274\mu$eV. We therefore conclude that the shift of 106 $\mu$eV observed in Fig.2a for $\sigma^+$ excitation corresponds to a degree of nuclear spin polarization $S_N=39\%$. This value is similar to the maximum degree of polarization obtained for InGaAs dots at low T \cite{Braun2,Tartakovskii}. 

The importance of the feedback mechanism observed in Fig.2a becomes less significant if a higher external field is applied since it becomes more difficult to compensate the external field. This is demonstrated in Fig.2b, where power dependences of the trion splitting $E_{xZ}$ at 5T are plotted. The maximum Overhauser shift observed for $\sigma^+$ excitation is 68 $\mu$eV, compared to 106 $\mu$eV at $B_{Z}=1.5$T. Although the nuclear spin pumping is still 
more efficient for $\sigma^+$ excitation, the process is markedly slowed down by the high external field, and is less sensitive to $B_N$ since $B_{Z}\gg B_N$.

The inset in Fig.2b also shows that a significant nuclear polarization can be optically pumped at zero external field. Trion splittings up to 28$\mu$eV are observed occurring solely due to the Overhauser field $B_N$, which can be estimated to be 0.3T with the corresponding degree of nuclear spin polarization $S_N\approx$10$\%$. These magnitudes of $B_N$ and $S_N$ are similar to those reported for the electron charged InGaAs dots in Ref.\cite{Lai}. Note that at $B_{Z}=0$ the pumping efficiencies are very similar for both $\sigma^+$ and $\sigma^-$ excitation.

Observation of nuclear polarization at zero field is another piece of evidence implying that we deal with negatively charged dots as has been observed in Ref.\cite{Lai}. We have performed similar experiments at $B=0$ on positively charged and neutral self-assembled InGaAs/GaAs and interface GaAs/AlGaAs. Nuclear spin pumping have not been observed. On the other hand, in negatively charged dots, the optically orientated resident electron can relax its spin due to the hyperfine interaction \cite{Lai,Eble}, thus leading to a build-up of nuclear spin polarization.  \\

\begin{figure}
\centering
\includegraphics[height=5cm]{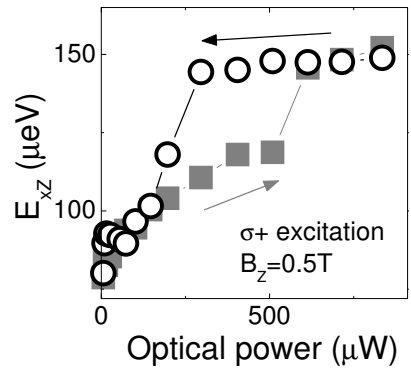}
\caption{Power dependence of the splitting $E_{xZ}$ (as defined in Fig.1d) measured for a single InP dot trion at $B_{Z}=0.5$T for $\sigma^+$ circularly polarized excitation. Arrows show the direction in which the power was scanned.}
\label{fig3}
\end{figure}

{\bf V. NUCLEAR SPIN BISTABILITY}

Positive feedback of $B_N$ on the spin flip-flop probability $p_{hf}$, as observed in Fig.2a, has been shown to lead to nuclear spin bi-stability in InGaAs/GaAs dots \cite{Braun2,Maletinsky1,Tartakovskii,Russell,Makhonin}. We find a similar behavior in negatively charged InP dots studied here.  However, we find that such bi-stability effects can only be observed at low magnetic fields. Fig.3 shows an example of the bi-stable behavior measured for a single InP dot at $B_{Z}=0.5$T. In this graph $E_{xZ}$ splitting is plotted as a function of power for $\sigma^+$ polarized excitation. When the power, $P$, is scanned from $P\approx0$, the exciton Zeeman splitting changes gradually from 80 to 118 $\mu$eV (shown in gray) up to $P\approx$500$\mu$W, where $E_{xZ}$ changes abruptly from 118 to 146 $\mu$eV. For higher powers $E_{xZ}$ shows very weak dependence on the excitation power.

The abrupt switching occurs when the Overhauser field has almost compensated $B_Z$, which will lead to a fast electron-nuclear spin transfer rate. 
If the power is now scanned from $P>1000\mu$W to zero, $E_{xZ}$ first shows a negligible dependence on power in the range $300<P<1000\mu$W. At $P\approx300\mu$W a very sharp decrease of $E_{xZ}$ is observed corresponding to the reduction of the nuclear polarization. In the range of powers  $200<P<600\mu$eV two stable nuclear spin states differing by $\Delta S_N\approx10\%$ are observed on the dot, constituting the observation of nuclear spin bi-stability.  

In general the observation of the optically induced bi-stability of the nuclear spin in a quantum dot is strongly dependent on the electron spin dynamics, determined in turn by the population and spin dynamics of all charge carriers on the dot~\cite{Russell}. On the other hand, it is possible to predict the range of external magnetic fields where the switching behavior as in Fig.3 can be observed. This will occur if the optically pumped Overhauser field can completely compensate the external field~\cite{Tartakovskii,Russell}. Although this is not the only necessary condition for the observation of the nuclear spin switch~\cite{GaAs}, this is a reliable starting criterion. The largest $B_N$ observed for the dots studied in this work is below 1.5T in contrast to InGaAs dots, where the Overhauser fields up to 3T have been observed~\cite{Braun2,Tartakovskii}. This indicates that in InP dots the occurrence of the nuclear spin bi-stability should be expected at low external magnetic fields as found in our  experiments where bi-stability is observed in the range of $B_{Z}\approx 0.3-1$T.\\

{\bf VI. SUMMARY}

In conclusion, strong nuclear spin effects are reported in individual optically pumped electron-charged InP dots due to dynamic nuclear polarization. This opens up the possibility to optically manipulate a polarized system of phosphorus nuclei, possessing the simplest nuclear spin configuration with $I_P=\pm1/2$ in a semiconductor nano-structure. The InP dots contain a common element, In, with the widely studied InGaAs/GaAs system and a comparative analysis of the two types of dots  is presented here. Similarly to InGaAs dots, the nuclear spin polarization on the dot is shown to produce a strong feedback on the nuclear spin pumping efficiency. A degree of polarization of $\approx40\%$ has been pumped optically, a limit very similar to InGaAs dots at low temperature. This limit might be related to a finite re-excitation rate of the dot, limiting supply of the electron spin to the dot, and low probability of the spin flip-flop. Development of nuclear magnetic resonance techniques is likely to shed further light on the spin transfer mechanism in nano-structures built from these complex semiconductor alloys. Electron-doped III-V QDs present an interesting optically controlled system of coupled electron and nuclear spins. Due to extremely long electron life-times on the dot (at low temperatures), it should be possible to achieve a regime where nuclear polarization becomes frozen with its decay suppressed due to the inhomogeneous Knight field induced by the localized electron.

This work has been supported by the Sheffield EPSRC Programme grant GR/S76076, the EPSRC IRC for Quantum Information Processing, ESF-EPSRC network EP/D062918 and by the Royal Society. AIT was supported by the EPSRC (grants EP/C54563X/1 and EP/C545648/1).


\begin{thebibliography}{}
\bibitem{GaAs} We find that in neutral interface GaAs/AlGaAs dots a complete compensation of the external field by $B_N$ is possible. However, the bi-stability has not so far been observed.
\end{thebibliography}
\end{document}